\documentclass[prb,twocolumn,floatfix]{revtex4}
\usepackage{graphicx}
\usepackage{color}
\usepackage[utf8]{inputenc}

\begin{document}
\title{
Half-filled Hubbard Model on a Bethe lattice with next-nearest neighbor hopping}
\author{Robert Peters}
\affiliation{Institute for Theoretical Physics, University of G\"ottingen, 
Friedrich-Hund-Platz 1,
37077 G\"ottingen, Germany}
\author{Thomas Pruschke}
\affiliation{Institute for Theoretical Physics, University of G\"ottingen, 
Friedrich-Hund-Platz 1,
37077 G\"ottingen, Germany}
\begin{abstract}
 We study the interplay between N\'eel-antiferromagnetism and the
 paramagnetic metal-insulator-transition (PMIT) on a
 Bethe lattice with nearest and next-nearest neighbor
 hopping $t_1$ and $t_2$. We  concentrate in this paper on the situation at
 half-filling. For $t_2/t_1\rightarrow 1$ the PMIT outgrows the
 antiferromagnetic phase and shows a scenario similar to V$_2$O$_3$. In this
 parameter regime we also observe a novel magnetic phase.
\end{abstract}
\pacs{}
\maketitle              
\section{Introduction}
Understanding correlation effects is one major goal of condensed
matter physics. Strong correlations manifest themselves in various
forms. The paramagnetic
Mott-Hubbard
metal-insulator-transition (PMIT)\cite{imada1998} is a
well-known 
and interesting example. With increasing interaction strength the
Fermi liquid state 
breaks down at a critical value and an
insulator is formed.\par Another fundamental example is magnetism, where
electrons reduce the energetic cost of the Coulomb-interaction
by ordering.
Both effects can of course occur simultaneously and are  the heart of
the extremely rich phase diagram of e.g. transition
metal compounds like for example V$_2$O$_3$ or
LaTiO$_3$\cite{mcwhan1973,katsufuji1997}. \par
Besides strong correlations, another major ingredient for the
understanding of the phase diagram of compounds like V$_2$O$_3$
is frustration. V$_2$O$_3$ crystallizes in the corundum
structure with the V-ions located on a honeycomb
lattice in the ab-plane, while along the c-axis a more 
complicated coordination is observed, which induces frustration of the 
magnetic interactions \cite{imada1998}.
Nevertheless does the phase
diagram of V$_2$O$_3$ show an antiferromagnetic phase at 
temperatures below $T_N\approx 180K$. Upon doping with Ti one may
suppress this order. Such a doping with a smaller ion can be viewed as internal
pressure,\cite{imada1998} hence the suppression of the magnetic order
is commonly 
interpreted as happening through an increase of the bandwidth
respectively a decrease of the correlation effects. Consequently,
the critical Ti doping is conventionally related to the existence of a
lower critical 
value of the 
electronic interaction parameter. At higher
temperatures the antiferromagnetic state becomes unstable towards a
paramagnet and one can eventually 
observe a paramagnetic metal-insulator-transition up to temperatures
$T\approx 400K$.\par 
Frustration is a quite common feature in real materials. 
Very interesting examples for 
frustrated systems are layered organic compounds like
$\kappa$-(BEDT-TTF)2X\cite{lefebvre2000,tsai2001,morita2002,kurosaki2005,sasaki2005,aryanpour2006,yokoyama2006,kyung2006,watanabe2006,koretsune2007,watanabe2008,ohashi2008,sasaki2008,nevidomskyy2008}. They have a similar
phase diagram as the high-temperature
superconductors (HTSC)\cite{mckenzie1997}. The 
phases of these organic systems are controlled by pressure and
frustration rather than by doping as in HTSC\cite{lee2006}. They are
usually described by an anisotropic triangular lattice, and
changing the anion (X) in
these systems modifies the frustration of the lattice. Besides
superconductivity also magnetic ordering and a PMIT can be found.\par 
These two examples by no means exhaust the zoo of materials showing
such interplay or competition between PMIT and ordered
phases\cite{imada1998}. For example, rare-earth compounds like
Ce(Rh,Ir)$_{1-x}$(Co,Ir)$_x$In$_5$ do show a similarly bizarre phase
diagram\cite{hegger2000}.
Besides their usually complicated lattice structure another challenge
for a theoretical description of such compounds is that the presence
of elements with partially filled d- or f-shells in principle requires
a multi-orbital description to account for effects like Hund's or
spin-orbit coupling properly. Furthermore the residual degeneracies in
the solid state crystalline environment lead to degenerate multiplets
which in turn can give rise to even more complex structures like
orbital order or polaron formation (see e.g. \textcite{imada1998} for
an overview and references).\par 
Although all these ingredients play an important role for a
quantitative theoretical description of transition-metal or rare-earth
compounds, we here want to focus on the one-orbital situation,
in particular on the relation between PMIT and
antiferromagnetism. This restriction to a simpler but by no means
trivial situation will enable us to investigate the relation between
these two paradigms of correlation effects with a small and
controllable set of parameters and thus obtain some hint to how both
phases interact. A model suitable for analyzing this kind of physics is
provided through the Hubbard
model\cite{hubbard1963,kanamori1963,gutzwiller1963} 
\begin{equation}
H=\sum_{i,j,\sigma}t_{ij}c^\dagger_{i\sigma}c_{j\sigma}+U\sum_in_{i\uparrow}n_{i\downarrow},
\end{equation}
where $c_{i\sigma}^\dagger(c_{i\sigma})$ creates (annihilates) an
electron with spin $\sigma$ at site $i$ and
$n_{i\uparrow}(n_{i\downarrow})$ is the density operator for spin
up(down) at site i. The parameters $t_{ij}$ represent the hopping
amplitude from 
$i$ to $j$ and $U$ is the interaction strength. In this paper we will
measure the interaction relative to the bandwidth, which is related to the
hopping amplitude. Although at first sight very simplistic, this model
is highly nontrivial. Besides other 
methods, especially in one dimension, progress in
understanding its physics was achieved by the
development of 
the dynamical mean field theory (DMFT)\cite{georges1996}. 
The DMFT is a very powerful tool for analyzing strongly correlated lattice
systems, mapping the lattice problem onto a quantum impurity problem,
which has to be solved self consistently. For solving this impurity
problem for arbitrary interaction strengths and temperatures we here use
Wilson's numerical 
renormalization group\cite{wilson1975,bulla2008}. An interesting fact
is that the only information about the lattice structure, which enters a
DMFT self consistency calculation, is the local density of
states (DOS) of the non-interacting system. We performed our calculations for a Bethe lattice with
nearest-neighbor (nn) and next-nearest-neighbor (nnn) hopping $t_1$
and $t_2$, respectively. 
\begin{figure}[htb]
\begin{center}
\includegraphics[width=0.45\textwidth,clip]{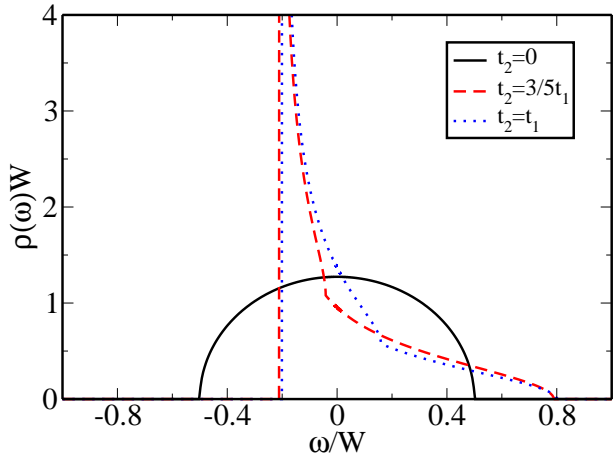}
\end{center}
\caption[]{(color online) DOS for increasing nnn-hopping
  $t_2$. Observe the van-Hove singularity at the lower band
  edge. $\rho(\omega)$ and $\omega$ are scaled with the bandwidth $W=4t_2+2t_1+t_1^2/(4t_2)\quad (t_2>1/4t_1)$.}\label{frustDOS}
\end{figure}
The DOS in this case can be calculated using a topological
ansatz\cite{kollar2005,eckstein2005}. Starting from a particle-hole
symmetric DOS at $t_2=0$ the density of states becomes now
asymmetric with increasing $t_2$ (see Fig. \ref{frustDOS}) and develops a van-Hove singularity at the lower band edge
for positive and increasing 
nnn-hopping $t_2$. In contrast to $t_2=0$, where the particle-hole symmetry
can be employed to fix the filling at $\langle n\rangle=1$ precisely,
the asymmetry present for $t_2\ne 0$ makes it more difficult to
perform calculations with the filling kept at $\langle n\rangle=1$
with sufficient accuracy. Thus DMFT calculations typically take much longer
here due to the necessary adjustment of the chemical potential. Of
course the Bethe lattice does not represent a lattice realized in real
materials. However, in contrast to the hypercubic 
lattice with infinite coordination number the Bethe lattice has a
compact support and thus  possesses band edges, which provides a more realistic
scenario. 
\par
Since the early days of DMFT, there have been many contributions by
different groups to the subject of the PMIT and
antiferromagnetism.\cite{pruschke2005,georges1996}
However, frustration effects up to now where introduced in DMFT typically
within the so-called two-sublattice fully frustrated
model\cite{rozenberg1995,georges1996,duffy1997,hofstetter1998,chitra1999,zitzler2004},
which results in a particle-hole symmetric DOS even with frustration. As side effect,
this way of introducing frustration leaves the paramagnetic phase
unchanged. For the non-frustrated system the PMIT
is then completely covered by the antiferromagnetic phase, which exists for
half-filling for all finite values of
$U$\cite{pruschke2005,dongen1991}.
For the frustrated system, on the other hand,
there exists a lower critical value for the interaction $U$, which
increases with increasing frustration. It was furthermore found that the
N\'{e}el-temperature decreases with increasing frustration such that the PMIT outgrows the
antiferromagnetic phase\cite{zitzler2004}. In early calculations using
this way of introducing
frustration based on exact diagonalization studies of the
two-sublattice fully frustrated
model\cite{georges1996,chitra1999,hofstetter1998}, the authors also found 
parameter regions in the phase diagram where an antiferromagnetic
metal appeared to be stable. However, this antiferromagnetic metal phase
was later traced back to numerical subtleties in the exact
diagonalization procedure and shown to be actually absent from the
phase diagram\cite{zitzler2004}.\par 
The first attempt to study the Hubbard model on the Bethe lattice with
correct inclusion of nn- and nnn-hopping has been performed rather
recently\cite{eckstein2007}. In this work the authors concentrated on the 
paramagnetic PMIT and found 
phase-separation between the insulating and metallic phase.
\par
In this paper we investigate the PMIT as well as
antiferromagnetism and concentrate on the competition between the
paramagnetic phase including the PMIT and the antiferromagnetic phase
at intermediate and high grades of frustration. We especially look at
the case $t_2\rightarrow t_1$ and raise the question, if the scenario of
the outgrowing PMIT, proposed in \textcite{zitzler2004}, still holds for
the correct asymmetric density of states. 
The paper is arranged as following. After this introduction we start
with a brief look at the PMIT, followed by a discussion of the phase diagram at half-filling including
antiferromagnetism and the PMIT. The
next paragraph addresses especially the case of very strong
frustration and the question how the magnetic order is realized
there. The paper will be closed  by a summary of our results and an outlook.  \par
\section{Metal-insulator-transition}
The metal insulator transition for the Bethe lattice with nnn-hopping
has been analyzed by \textcite{eckstein2007} within the self-energy
functional approach\cite{potthoff2003}. They particularly focused
on $t_2/t_1=3/7$ and discussed an unexpected occurrence of phase
separation in the paramagnetic state between a Mott-Hubbard insulator
and a correlated metal at and near half-filling.
Here we want to investigate the behavior of the system as function of
increasing frustration. Due to
symmetry there is no difference between $t_2$ and $-t_2$. The
calculations were done using Wilson's NRG as impurity solver for the
DMFT, with $\Lambda=2$, $1800$ states kept per NRG step and a
logarithmic broadening $b=0.8$ to obtain spectral functions.
We want to note at this point that the choice of NRG numerical
parameters does not influence the qualitative nature of the results. We
observe, however, small dependencies on $\Lambda$ and $b$, which tend to
become more pronounced close to phase transitions and may result in
systematic errors in numerical values for critical parameters 
of the order of $<5\%$.\cite{bulla2008} 

\begin{figure}[htb]
\begin{center}
\includegraphics[width=0.45\textwidth,clip]{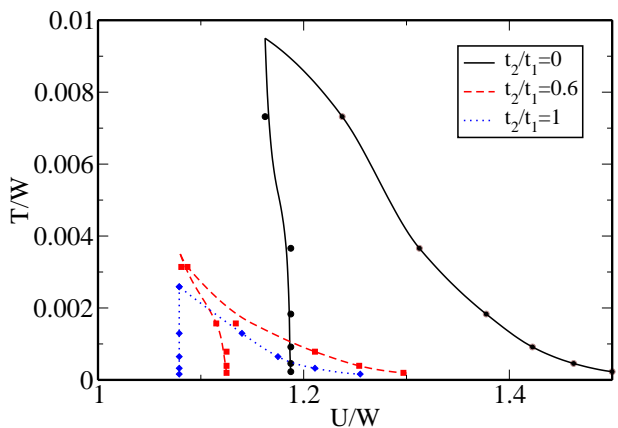}
\end{center}
\caption[]{(color online) The transition lines for the PMIT for
  different frustrations as function of temperature and interaction
  strength. For each 
  frustration the right line represents the transition from the metal
  to the insulator, while the left line represents the transition from
  the insulator to the metal. Symbols mark the calculated data points,
the lines are fits meant as guide to the eye.}\label{PMIT}
\end{figure}

Figure
\ref{PMIT} shows the paramagnetic metal insulator transition for
various values $t_2/t_1$. As energy-scale we choose the bandwidth
$$
W=\left\{\begin{array}{l}\mbox{$4t_1$ for $0\le \vert t_2\vert \le t_1/4$}\\[5mm]
    \mbox{$4\vert t_2\vert +2t_1+t_1^2/(4\vert t_2\vert)$ for $\vert t_2\vert >t_1/4$}
  \end{array}\right.
$$

of the non-interacting system. 
Note that these results are obtained by artificially suppressing an 
antiferromagnetic instability. We will come back to this point
later. The occupation was kept fixed at $n=1\pm 0.005$ by adjusting the chemical
potential. Note that in contrast to the case with $t_2=0$ it is not
possible to achieve $n=1$ here within numerical precision  due to the asymmetric DOS (see Fig. \ref{frustDOS}).
For increasing $t_2/t_1\rightarrow 1$ the PMIT is shifted towards lower
interaction strengths and also lower temperatures. While the shift in
the interaction strength is rather moderate, we notice a large
difference in the temperature of the critical endpoint between the
unfrustrated and highly frustrated system. This observation of course renews our interest in the question, to what extent long-range hopping can help to push the paramagnetic MIT out of the expected antiferromagnetic phase for reasonable magnitudes of $t_2$ to create a phase diagram similar to the one found for V$_2$O$_3$. The scenario proposed by
\textcite{zitzler2004} relied on the fact that the paramagnetic phase
 largely remains unaltered with increasing $t_2$. As the N\'eel-temperature for the antiferromagnet is reduced at the same time, the
 PMIT can eventually outgrow the 
antiferromagnetic phase.\par
\section{Antiferromagnetism at finite $t_2$}
\begin{figure}[htb]
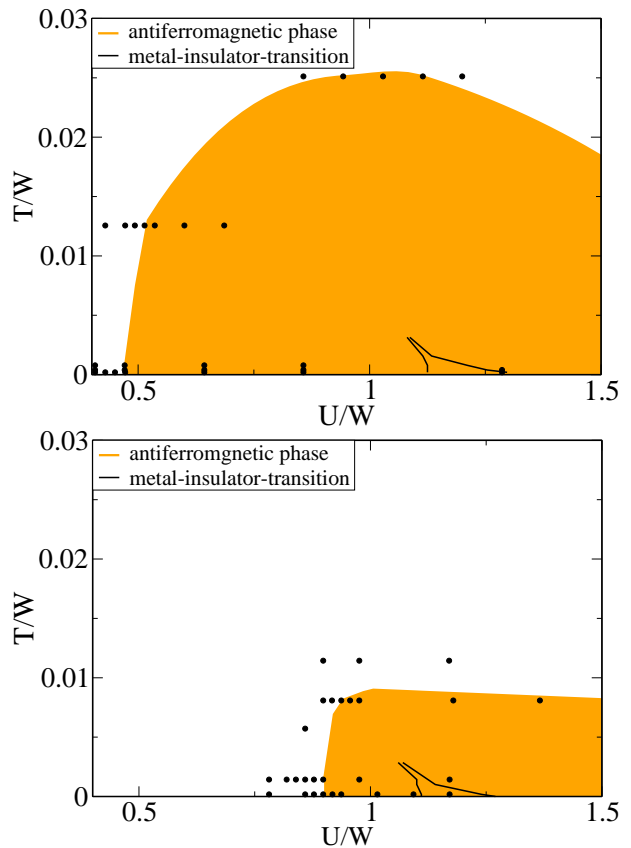

\begin{center}
\includegraphics[width=0.45\textwidth,clip]{phaset2_003}
\includegraphics[width=0.45\textwidth,clip]{phaset2_004}
\end{center}
\caption[]{(color online) The upper (lower) panel shows the $T-U$ phase
  diagram for $t_2/t_1=0.6$ $(0.8)$. The shaded area represents the
  antiferromagnetic phase, while the white area represents the
  paramagnetic phase. The lines show, where the PMIT in the
  paramagnetic phase would occur. The points denote the parameter values,
  where DMFT calculations were done. From these points
    the shaded area was constructed as guide to the eye. Additional calculations were
  performed to find the PMIT-lines.} \label{phase1}
\end{figure}

We now allow for antiferromagnetic ordering in our calculations. To
this end we reformulate the DMFT for an AB lattice
structure\cite{georges1996,pruschke2005} to accommodate the N\'eel ordering
and initialize the calculation with a small staggered field, which is
turned off after one DMFT iteration. The system then either
evolves into a paramagnetic or antiferromagnetic state with increasing
number of DMFT iterations.  Figure
\ref{phase1} shows the resulting phase diagrams for $t_2/t_1=0.6$
(upper panel) and
$t_2/t_1=0.8$ (lower panel) for different temperatures and interaction
strengths. 
The small black points show the locations, where calculations have
actually been performed. From
these data the shaded areas were constructed representing the 
antiferromagnetic phases. This of course means that the
phase boundaries shown here must be considered as guess only. However,
as we do not expect any strange structures to appear, this guess will presumably
represent the true phase boundary within a few percent.

The full lines in Fig.\ \ref{phase1} are the PMIT
transitions. Note that for both diagrams the same division of axes
was chosen. \par
In contrast to the Hubbard model on a bipartite lattice with $t_2=0$,
there now exists a finite critical value $U_c^{AF}$, below which no
antiferromagnetism can be stabilized even for temperature $T\to 0$.  
With increasing frustration the paramagnetic-antiferromagnetic
transition is shifted 
towards higher interaction strengths and lower temperatures, while the
PMIT is shifted towards lower interactions strengths. So obviously the
PMIT is shifted towards the phase boundaries of the antiferromagnetic
dome. So far this is the expected effect of the nnn-hopping  which
introduces frustration to the antiferromagnetic exchange. However,
note that although $t_2/t_1=0.8$ represents already a very strongly
frustrated 
system, the PMIT still lies well covered within the antiferromagnetic
phase. \par
Let us now have a closer look at the 
paramagnetic-antiferromagnetic transition.
Here, \textcite{zitzler2004} made the prediction that one has to
expect a first order transition close to the critical $U_c^{AF}$ at low
temperatures; while at larger values of $U$ again a second order
transition was found.\par
\begin{figure}[htb]
\begin{center}
\includegraphics[width=0.45\textwidth,clip]{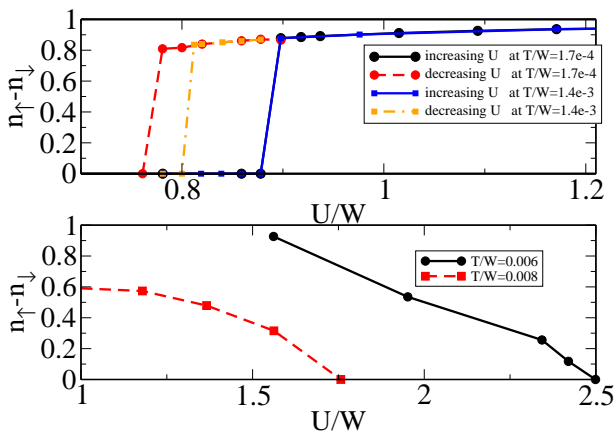}
\end{center}
\caption[]{(color online) Staggered
magnetization versus interaction $U/W$ for two different
temperatures and $t_2/t_1=0.8$. In the upper panel there are for each temperature two
transition lines, representing either increasing or decreasing
interaction strength. The region between both lines embodies a hysteresis
region. The lower panel shows the transition for large
interaction strengths. Here no hysteresis region could be found, but a
smooth transition.} \label{hysteresis2}  
\end{figure}
\begin{figure}[htb]
\begin{center}
\includegraphics[width=0.45\textwidth,clip]{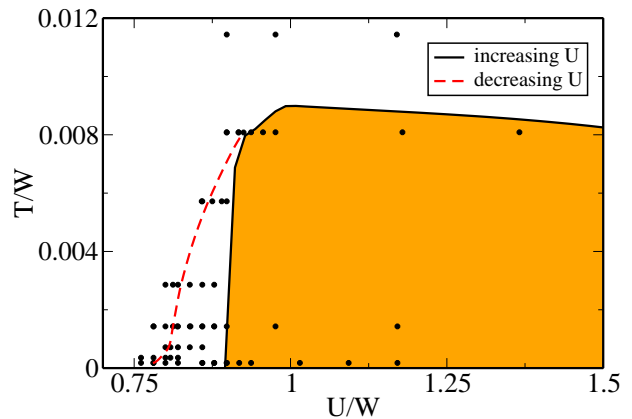}
\end{center}
\caption[]{(color online) Phase diagram for $t_2/t_1=0.8$
  including the temperature depending hysteresis region (between dashed
  and full line). The meaning of symbols is as in Fig. \ref{phase1}. The shaded area is as before meant as guide to the eye. } \label{hysteresis}
\end{figure}

 Figure \ref{hysteresis2}
shows the staggered magnetization for different temperatures
and interaction strengths at fixed $t_2/t_1=0.8$. The upper panel collects data for the
transition at low 
temperatures at the lower edge of the antiferromagnetic phase. The
full lines represent the transition from the paramagnetic to the
antiferromagnetic state with increasing interaction strength for two
different temperatures, while the dashed 
lines represent the transitions from the antiferromagnetic to the
paramagnetic state with decreasing interaction strength. In the upper panel
(small $U$)
one can clearly see a hysteresis of the antiferromagnetic transition. This hysteresis as well as
the jump in the magnetization are clear signs for a first order
transition. This antiferromagnetic hysteresis is very pronounced for
strong frustration
but numerically not resolvable for example for $t_2/t_1=0.2$. We
believe that the hysteresis 
region shrinks with decreasing $t_2$  and eventually cannot be resolved
anymore with numerical techniques. The whole temperature depending
hysteresis region can be 
seen in Fig. \ref{hysteresis} for the case $t_2/t_1=0.8$. One can see clearly the shrinking of
the hysteresis region with increasing temperature. Note that such a
hysteresis is also found in the two-sublattice fully frustrated
model,\cite{zitzler2004} which means that this quite likely is a generic
effect in frustrated systems at intermediate coupling strengths.\par
The lower panel in Fig. \ref{hysteresis2} shows
the staggered magnetization for temperatures just below the
corresponding 
N\'{e}el-temperatures and at
large interaction strengths. Here the magnetization
vanishes smoothly, which is the behavior expected for a second
order phase transition. In summary we thus find a 
first order transition at the critical interaction $U_c^{AF}$ where antiferromagnetism
sets in, and a second order transition for the large Coulomb parameter $U\gg W$. 
The merging from both transition lines is an interesting point in
itself. There must be a critical point where the first order
transition changes into a second order transition. It is however not
possible to resolve this merging within DMFT/NRG. First, the
logarithmic discretization of the temperatures within the NRG does not
allow to resolve this merging-region with arbitrary precision. Second,
the magnetization 
of the system becomes very small in this region, so it is not possible 
to distinguish between a (tiny) jump and numerical artifacts of a smoothly
vanishing order parameter. Consequently, we cannot decide anymore of
what order 
the transition will be. \par
Antiferromagnetic metallic phases at half-filling were reported in
earlier publications\cite{rozenberg1995, 
  duffy1997,hofstetter1998,chitra1999}. In our calculations we saw no
 evidence for an antiferromagnetic metallic state at
half-filling. Especially for strong frustration
$t_2/t_1\approx0.8$ the system directly jumps from a paramagnetic
metallic solution into an antiferromagnetic insulating solution with
high magnetization. In the papers cited, the region showing an
antiferromagnetic metallic solution broadens with increasing
$t_2$. This prediction we clearly cannot confirm, as discussed
above. Only in systems with 
small to intermediate frustration there are narrow
interaction regimes where we observe a small finite weight at the
Fermi level. One must however consider that the occupation number is
not exactly one but only 
within $0.5\%$. Also it was sometimes difficult to stabilize a
DMFT solution in these regions. In summary, we cannot see any clear
signs for 
an antiferromagnetic metallic state at half-filling in our
calculations. If any exists, then only for rather low frustration in a
very small regime about the critical interaction. To what extent these
rather special conditions can then be considered as realistic for real
materials is yet another question.  
\section{Nearly fully frustrated system}
In this last paragraph we want to study the situation, in which $t_1$
and $t_2$ are comparable in strength. 
Interestingly, there has been no attempt to calculate the phase
diagram on a mean-field level in the strongly frustrated model
$t_2\approx t_1$.
Therefore, before discussing the results of the DMFT calculations for
strongly frustrated systems $t_2/t_1\approx 1$ let us try to gain some
insight into the physics we must expect, by inspecting 
 classical spins on a Bethe lattice with
nn-interaction $J_1$ and nnn-interaction $J_2$.
Allowing that 
nearest-neighbor spins enclose an angle $\theta$ one ends up with
the  energy functional
\begin{eqnarray}
E/2N&=&J_1Z\cos(\theta)\nonumber\\
&+&J_2Z\sum_{i=1}^{Z-1}\left(\cos(\theta)^2
+\sin(\theta)^2\cos(2\pi
  i/Z)\right)
\end{eqnarray}
Performing the same limits and scaling as in DMFT one finds (see
Appendix A)
\begin{equation}
E/2N=J_1\cos(\theta)+J_2\cos(\theta)^2.
\end{equation}
Thus, the N\'{e}el-state with $\theta=\pi$ is the stable ground state
for 
$J_2<\frac{1}{2}J_1$, while one finds a spin wave with
$\theta=\pi-\arccos(J_1/2J_2)$ for
$J_2>\frac{1}{2}J_1$. \par
For the DMFT calculations 
we can allow only for solutions commensurate with the lattice. This
however will possibly be inconsistent with the spin structure favored
by the system. If, for 
example, we perform a
calculation focusing on the ferromagnetic solution within a parameter
regime, where the system 
wants to order antiferromagnetically, DMFT will not converge.
To investigate spin wave states with periodicities with more than two
lattice sites, one has to set up the correct DMFT self-consistency equations
respecting the lattice structure. While for a system on an infinite
Bethe lattice with nn  
hopping only it is straightforward to extend 
the DMFT to commensurate magnetic structures with periodicities of
more than two lattice sites, we did not
succeed in devising a scheme 
that allows for such calculations for systems with nnn hopping. The
reason is that one has to partition the lattice into an $ABCD\ldots$
structure. However, the nnn hopping makes it impossible to uniquely
identify the connectivity of the respective sublattices. A method
proposed by \textcite{fleck1999} for the two-dimensional cubic lattice is not
applicable in our case.
\par
\begin{figure}[htb]
\begin{center}
\includegraphics[width=0.45\textwidth,clip]{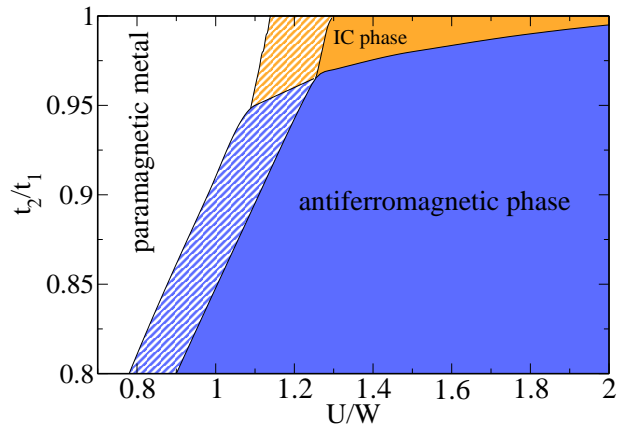}
\end{center}
\caption[]{(color online) Ground state ($T=0$) phase diagram for
  different strengths of frustration as function of the
  interaction. The 
  brindled regions are hysteresis regions for increasing or decreasing
interaction. The phase boundaries of the incommensurate (IC) phase
are only qualitative (explanation see text).} \label{hight2_U}
\end{figure}

\begin{figure}[htb]
\begin{center}
\includegraphics[width=0.45\textwidth,clip]{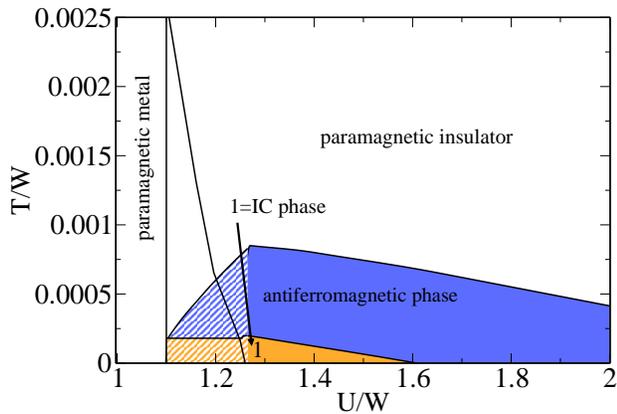}
\end{center}
\caption[]{(color online) Phase diagram $T-U$ for $t_2/t_1=0.98$
  including the PMIT. The brindled area represents again the
  hysteresis region.} \label{hight2_T}
\end{figure}

\begin{figure}[htb]
\begin{center}
\includegraphics[width=0.45\textwidth,clip]{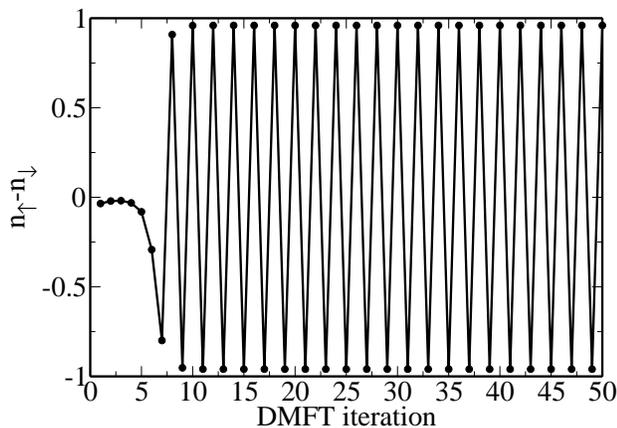}
\end{center}
\caption[]{Example of a non-convergent DMFT calculation. The
figure shows
the staggered polarization over the DMFT iteration number for $t_2/t_1=0.98$,
$U/W=1.35$ and $T\approx 0$. The lines are meant as guide to the eye.} \label{noncon}
\end{figure}

 
We thus only allowed for paramagnetic, ferromagnetic and
antiferromagnetic 
solutions  in our calculations. The resulting phase diagrams for
$t_2\rightarrow 
t_1$ are shown in 
Figs. \ref{hight2_U} and \ref{hight2_T}. Figure \ref{hight2_U}
displays the  
ground states for different grades of frustration and interaction
strengths. For $t_2/t_1<0.95$ the phase diagram has the same structure
as for small and intermediate $t_2$. The critical interaction strength 
$U_c^{AF}$ necessary to stabilize the N\'{e}el-state increases and for all
values above 
$U_c^{AF}$ we find an antiferromagnetic phase with a hysteresis region at 
the phase boundary. 
For $0.95<t_2/t_1<1$ the critical value $U_c^{AF}$ one needs to stabilize
the N\'{e}el-state 
increases dramatically. For $t_2=t_1$ finally we do not find an
antiferromagnetic 
N\'{e}el-state at all for any interaction strength $U$. Our DMFT
calculations however indicate that in this range of $t_2/t_1$ there
actually does  exists another magnetic
phase. Namely, for sufficiently small temperatures one obtains a
finite spin polarization in every DMFT iteration. However, the DMFT does
not converge to a unique state as function of DMFT iterations (see also Fig. \ref{noncon}).
In the
phase diagrams in Figs. \ref{hight2_U} and \ref{hight2_T} we have
named this regime the incommensurate phase (IC phase). In
this parameter regime the N\'eel-state becomes unstable towards the  behaviour
shown in Fig.\ \ref{noncon}. Here one can switch between a conventional N\'eel-state and the IC phase by only a small change of the interaction strength.
Note, that the phase
boundaries shown in the figure must be taken with some care as we
cannot compare the energies of  
the N\'{e}el-state and this IC phase 
to properly determine the phase boundaries. As we observe precisely the
same behavior for all investigated values   
$t_2/t_1=\{0.96,0.97,0.98,0.99,1.0\}$ we are convinced that the ground
state in this region is an incommensurate state, as to be expected from
our results for $S=\infty$. Similar observations also hold for
finite temperatures as shown in
Fig. \ref{hight2_T}, where the $T-U$ phase diagram for fixed
$t_2/t_1=0.98$ is displayed. For increasing interactions and $T=0$
there first is a 
transition from a paramagnetic metal to the IC phase and for
$U/W\approx 1.6$ from the IC phase to the N\'{e}el-state. For
increasing temperature the IC phase eventually becomes
unstable towards  
the N\'{e}el-state. In Fig. \ref{hight2_T} one can also see the PMIT
lines. As one can see it lies within the hysteresis region of the
magnetic phases but clearly outgrows both magnetic phases. This is the
scenario described in \textcite{zitzler2004}. 
\section{summary}
We studied the DMFT phase diagram of the Hubbard model at half-filling 
in the presence of nn- and nnn-hopping. 
In contrast to previous investigations we did our calculations for a
Bethe lattice with proper nnn hopping $t_2$, introducing a highly
asymmetric DOS already for the non-interacting system.\par
The first important observation concerns the paramagnetic
metal-insulator transition, which is suppressed by increasing $t_2$,
but at the same time shifted to lower values of the Coulomb
interaction.\par 
The at $t_2=0$ ubiquitous antiferromagnetic phase on the other hand
is suppressed up to a critical value $U_c^{AF}(t_2)$ with increasing $t_2$,
as expected. Furthermore, a hysteresis region between the
paramagnetic metal at small U and the antiferromagnetic insulator at
large U develops, showing that the transition is of first order. Note
that we did not observe any evidence for an antiferromagnetic metal
close to the phase boundary, nor did the PMIT reach out of the
antiferromagnetic insulator up to values $t_2=0.8t_1$.\par 
Thus far the observations are similar to the results found by
\textcite{zitzler2004} for the two-sublattice fully frustrated Bethe
lattice\cite{georges1996}. The shift of the PMIT to lower values of
$U$ together with a 
moderate suppression of the critical temperature for larger $t_2$
however motivated a more detailed investigation of the region of
larger $t_2$. A simple argument based on classical spins with competing
interactions showed that one has to expect an additional incommensurate
phase here. 
In fact, as already anti\-cipated qualitatively by
\textcite{zitzler2004}, for frustrations $0.96<t_2/t_1<1.0$ we
eventually found that the PMIT 
lies within the hysteresis region of the antiferromagnetic phase for
$T=0$, but
outgrows it in temperature. For such strong frustration we also 
found evidence for another magnetic phase besides ferromagnetism or
antiferromagnetism. Unfortunately this phase could not be stabilized
within our DMFT calculations, so its real nature remains
open. In connection with our argument based on classical
spins we believe that we can interpret the observed structure as
an incommensurate phase. This conjecture is further supported by the
fact, that for $t_1=t_2$ we found no
antiferromagnetic solution of the N\'eel-type, but only this frustrated
magnetic phase.\par
Especially the latter findings make it highly desirable to set up a
scheme that allows to study commensurable structures with period
beyond N\'eel-type for arbitrary lattice structures including
longer-ranged hopping.  
\begin{acknowledgments}
We want to thank Martin Eckstein and Dr. Markus Kollar for many helpful
discussions on the Bethe lattice with nnn hopping, and Dr. Timo Aspelmeier
for his help with 
the vector-spins. This work was supported by the DFG through 
PR298/10. Computer support was provided by the 
Gesellschaft f\"ur 
wissenschaftliche Datenverarbeitung in G\"ottingen and the
Norddeutsche Verbund f\"ur Hoch- und H\"ochstleistungsrechnen.
\end{acknowledgments}
\begin{appendix}
\section{calculation for vector-spins}
Here we want to present the calculation for 3-dimensional vector spins
on a Bethe lattice with antiferromagnetic coupling between nn- and
nnn-lattice sites.
\begin{figure}[htb]
\begin{center}
\includegraphics[width=0.45\textwidth,clip]{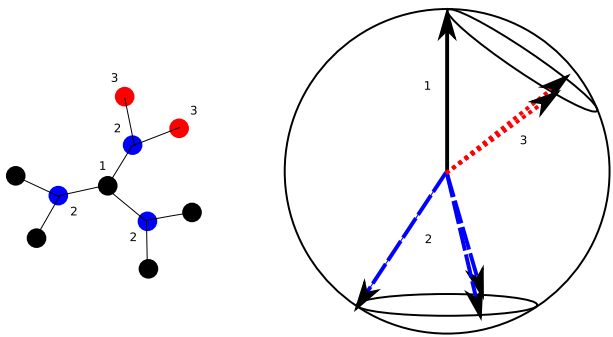}
\end{center}
\caption[]{(color online) Left: Bethe lattice $Z=3$ with
  sites numbered according to the vector spins on the
  right. The nearest neighbors of site 1 must lie on a circle so 
  there is an angle $\theta$ between $1$ and $2$. Similarly the
  nearest neighbors of one of the $2$-spins must lie on a circle
  including the $3$-spins and the $1$-spin.} \label{vectorspins}\end{figure}
We take $Z$ nearest neighbors and interaction strengths $J_1$
between nn-sites and $J_2$ between nnn-sites. The last parameter
entering this calculation is the angle $\theta$ between nn-spins.
Although the initial assumption that two neighboring spins form an 
angle theta may seem somewhat restrictive, we are not aware of other 
configurations with lower
total energy \cite{Aspelmeier}.
We want to 
minimize the energy with respect to this angle. According to
Fig. \ref{vectorspins} the 
nn-spins of one spin, must lie on a circle.
The spins ending on the circle are all nnn-spins. Due to the
antiferromagnetic interaction $J_2$, we assume that they want to
maximize the angle between them. Since there are $Z$ spins on each
circle, we assume they will have angle $2\pi/Z$ projected on the
circle. Using now simple trigonometry, the angle between nnn-spins is
given by 
\begin{eqnarray}
\cos{\gamma}&=&\frac{2-(2R^2-2R^2\cos(2\pi
  i/Z))}{2}\nonumber\\
&=&\cos(\theta)^2+\sin(\theta)^2\cos(2\pi i/Z),\nonumber
\end{eqnarray}
where $i$ runs from $0$ to $Z-1$, giving the different positions on one
circle. Inserting this into the Hamiltonian
\begin{displaymath}
E=J_1\sum_{i,j\in nn}\vec{S_i}\vec{S_j}+J_2\sum_{i,j\in
  nnn}\vec{S_i}\vec{S_j},
\end{displaymath}
one finds for the energy
\begin{eqnarray}
E/2N&=&J_1Z\cos(\theta)\nonumber\\
&+&J_2Z\sum_{i=1}^{Z-1}\left(\cos(\theta)^2+\sin(\theta)^2\cos(2\pi
  i/Z)\right).\nonumber
\end{eqnarray}
Performing now the limit $Z\rightarrow\infty$ and scaling
$J_1Z\rightarrow J_1^\star$ and $J_2Z\rightarrow J_2^\star/Z$ one
finally obtains for the energy per lattice site
\begin{displaymath}
E_{Z=\infty}(\theta)/(2N)=J_1^\star\cos(\theta)+J_2^\star\cos(\theta)^2.
\end{displaymath}
One can now see that the N\'{e}el-state $\theta=\pi$ is stable for
$J_2^\star/J_1^\star<1/2$, because $d^2E(\theta=\pi)/d\theta^2=J_1^\star-2J_2^\star$.
\end{appendix}


\begin{thebibliography}{38}
\expandafter\ifx\csname natexlab\endcsname\relax\def\natexlab#1{#1}\fi
\expandafter\ifx\csname bibnamefont\endcsname\relax
  \def\bibnamefont#1{#1}\fi
\expandafter\ifx\csname bibfnamefont\endcsname\relax
  \def\bibfnamefont#1{#1}\fi
\expandafter\ifx\csname citenamefont\endcsname\relax
  \def\citenamefont#1{#1}\fi
\expandafter\ifx\csname url\endcsname\relax
  \def\url#1{\texttt{#1}}\fi
\expandafter\ifx\csname urlprefix\endcsname\relax\def\urlprefix{URL }\fi
\providecommand{\bibinfo}[2]{#2}
\providecommand{\eprint}[2][]{\url{#2}}

\bibitem[{\citenamefont{Imada et~al.}(1998)\citenamefont{Imada, Fujimori, and
  Tokura}}]{imada1998}
\bibinfo{author}{\bibfnamefont{M.}~\bibnamefont{Imada}},
  \bibinfo{author}{\bibfnamefont{A.}~\bibnamefont{Fujimori}}, \bibnamefont{and}
  \bibinfo{author}{\bibfnamefont{Y.}~\bibnamefont{Tokura}},
  \bibinfo{journal}{Rev. Mod. Phys.} \textbf{\bibinfo{volume}{70}},
  \bibinfo{pages}{1039} (\bibinfo{year}{1998}).

\bibitem[{\citenamefont{McWhan et~al.}(1973)\citenamefont{McWhan, Remeika,
  Brinkman, and Rice}}]{mcwhan1973}
\bibinfo{author}{\bibfnamefont{D.~B.} \bibnamefont{McWhan}},
  \bibinfo{author}{\bibfnamefont{J.~P.} \bibnamefont{Remeika}},
  \bibinfo{author}{\bibfnamefont{W.~F.} \bibnamefont{Brinkman}},
  \bibnamefont{and} \bibinfo{author}{\bibfnamefont{T.~M.} \bibnamefont{Rice}},
  \bibinfo{journal}{Phys. Rev. B} \textbf{\bibinfo{volume}{7}},
  \bibinfo{pages}{1920} (\bibinfo{year}{1973}).

\bibitem[{\citenamefont{Katsufuji et~al.}(1997)\citenamefont{Katsufuji,
  Taguchi, and Tokura}}]{katsufuji1997}
\bibinfo{author}{\bibfnamefont{T.}~\bibnamefont{Katsufuji}},
  \bibinfo{author}{\bibfnamefont{Y.}~\bibnamefont{Taguchi}}, \bibnamefont{and}
  \bibinfo{author}{\bibfnamefont{Y.}~\bibnamefont{Tokura}},
  \bibinfo{journal}{Phys. Rev. B} \textbf{\bibinfo{volume}{56}},
  \bibinfo{pages}{10145} (\bibinfo{year}{1997}).

\bibitem[{\citenamefont{Lefebvre et~al.}(2000)\citenamefont{Lefebvre, Wzietek,
  Brown, Bourbonnais, Jérome, Mézière, Fourmigué, and
  Batail}}]{lefebvre2000}
\bibinfo{author}{\bibfnamefont{S.}~\bibnamefont{Lefebvre}},
  \bibinfo{author}{\bibfnamefont{P.}~\bibnamefont{Wzietek}},
  \bibinfo{author}{\bibfnamefont{S.}~\bibnamefont{Brown}},
  \bibinfo{author}{\bibfnamefont{C.}~\bibnamefont{Bourbonnais}},
  \bibinfo{author}{\bibfnamefont{D.}~\bibnamefont{Jérome}},
  \bibinfo{author}{\bibfnamefont{C.}~\bibnamefont{Mézière}},
  \bibinfo{author}{\bibfnamefont{M.}~\bibnamefont{Fourmigué}},
  \bibnamefont{and} \bibinfo{author}{\bibfnamefont{P.}~\bibnamefont{Batail}},
  \bibinfo{journal}{Phys. Rev. Lett.} \textbf{\bibinfo{volume}{85}},
  \bibinfo{pages}{5420} (\bibinfo{year}{2000}).

\bibitem[{\citenamefont{Tsai and Marston}(2001)}]{tsai2001}
\bibinfo{author}{\bibfnamefont{S.-W.} \bibnamefont{Tsai}} \bibnamefont{and}
  \bibinfo{author}{\bibfnamefont{J.}~\bibnamefont{Marston}},
  \bibinfo{journal}{Can. J. Phys.} \textbf{\bibinfo{volume}{79}},
  \bibinfo{pages}{1643} (\bibinfo{year}{2001}).

\bibitem[{\citenamefont{Morita et~al.}(2002)\citenamefont{Morita, Watanabe, and
  Imada}}]{morita2002}
\bibinfo{author}{\bibfnamefont{H.}~\bibnamefont{Morita}},
  \bibinfo{author}{\bibfnamefont{S.}~\bibnamefont{Watanabe}}, \bibnamefont{and}
  \bibinfo{author}{\bibfnamefont{M.}~\bibnamefont{Imada}}, \bibinfo{journal}{J.
  Phys. Soc. Jpn.} \textbf{\bibinfo{volume}{71}}, \bibinfo{pages}{2109}
  (\bibinfo{year}{2002}).

\bibitem[{\citenamefont{Kurosaki et~al.}(2005)\citenamefont{Kurosaki, Shimizu,
  Miyagawa, Kanoda, and Saito}}]{kurosaki2005}
\bibinfo{author}{\bibfnamefont{Y.}~\bibnamefont{Kurosaki}},
  \bibinfo{author}{\bibfnamefont{Y.}~\bibnamefont{Shimizu}},
  \bibinfo{author}{\bibfnamefont{K.}~\bibnamefont{Miyagawa}},
  \bibinfo{author}{\bibfnamefont{K.}~\bibnamefont{Kanoda}}, \bibnamefont{and}
  \bibinfo{author}{\bibfnamefont{G.}~\bibnamefont{Saito}},
  \bibinfo{journal}{Phys. Rev. Lett} \textbf{\bibinfo{volume}{95}},
  \bibinfo{pages}{177001} (\bibinfo{year}{2005}).

\bibitem[{\citenamefont{Sasaki et~al.}(2005)\citenamefont{Sasaki, Yoneyama,
  Suzuki, Kobayashi, Ikemoto, and Kimura}}]{sasaki2005}
\bibinfo{author}{\bibfnamefont{T.}~\bibnamefont{Sasaki}},
  \bibinfo{author}{\bibfnamefont{N.}~\bibnamefont{Yoneyama}},
  \bibinfo{author}{\bibfnamefont{A.}~\bibnamefont{Suzuki}},
  \bibinfo{author}{\bibfnamefont{N.}~\bibnamefont{Kobayashi}},
  \bibinfo{author}{\bibfnamefont{Y.}~\bibnamefont{Ikemoto}}, \bibnamefont{and}
  \bibinfo{author}{\bibfnamefont{H.}~\bibnamefont{Kimura}},
  \bibinfo{journal}{J. Phys. Soc. Jpn.} \textbf{\bibinfo{volume}{74}},
  \bibinfo{pages}{2351} (\bibinfo{year}{2005}).

\bibitem[{\citenamefont{Aryanpour et~al.}(2006)\citenamefont{Aryanpour,
  Pickett, and Scalettar}}]{aryanpour2006}
\bibinfo{author}{\bibfnamefont{K.}~\bibnamefont{Aryanpour}},
  \bibinfo{author}{\bibfnamefont{W.~E.} \bibnamefont{Pickett}},
  \bibnamefont{and} \bibinfo{author}{\bibfnamefont{R.~T.}
  \bibnamefont{Scalettar}}, \bibinfo{journal}{Phys. Rev. B}
  \textbf{\bibinfo{volume}{74}}, \bibinfo{pages}{085117}
  (\bibinfo{year}{2006}).

\bibitem[{\citenamefont{Yokoyama et~al.}(2006)\citenamefont{Yokoyama, Ogata1,
  and Tanaka}}]{yokoyama2006}
\bibinfo{author}{\bibfnamefont{H.}~\bibnamefont{Yokoyama}},
  \bibinfo{author}{\bibfnamefont{M.}~\bibnamefont{Ogata1}}, \bibnamefont{and}
  \bibinfo{author}{\bibfnamefont{Y.}~\bibnamefont{Tanaka}},
  \bibinfo{journal}{J. Phys. Soc. Jpn.} \textbf{\bibinfo{volume}{75}},
  \bibinfo{pages}{114706} (\bibinfo{year}{2006}).

\bibitem[{\citenamefont{Kyung and Tremblay}(2006)}]{kyung2006}
\bibinfo{author}{\bibfnamefont{B.}~\bibnamefont{Kyung}} \bibnamefont{and}
  \bibinfo{author}{\bibfnamefont{A.-M.~S.} \bibnamefont{Tremblay}},
  \bibinfo{journal}{Phys. Rev. Lett} \textbf{\bibinfo{volume}{97}},
  \bibinfo{pages}{046402} (\bibinfo{year}{2006}).

\bibitem[{\citenamefont{Watanabe et~al.}(2006)\citenamefont{Watanabe, Yokoyama,
  Tanaka, and ichiro Inoue}}]{watanabe2006}
\bibinfo{author}{\bibfnamefont{T.}~\bibnamefont{Watanabe}},
  \bibinfo{author}{\bibfnamefont{H.}~\bibnamefont{Yokoyama}},
  \bibinfo{author}{\bibfnamefont{Y.}~\bibnamefont{Tanaka}}, \bibnamefont{and}
  \bibinfo{author}{\bibfnamefont{J.}~\bibnamefont{ichiro Inoue}},
  \bibinfo{journal}{J. Phys. Soc. Jpn.} \textbf{\bibinfo{volume}{75}},
  \bibinfo{pages}{074707} (\bibinfo{year}{2006}).

\bibitem[{\citenamefont{Koretsune et~al.}(2007)\citenamefont{Koretsune, Motome,
  and Furusaki}}]{koretsune2007}
\bibinfo{author}{\bibfnamefont{T.}~\bibnamefont{Koretsune}},
  \bibinfo{author}{\bibfnamefont{Y.}~\bibnamefont{Motome}}, \bibnamefont{and}
  \bibinfo{author}{\bibfnamefont{A.}~\bibnamefont{Furusaki}},
  \bibinfo{journal}{J. Phys. Soc. Jpn.} \textbf{\bibinfo{volume}{76}},
  \bibinfo{pages}{074719} (\bibinfo{year}{2007}).

\bibitem[{\citenamefont{Watanabe et~al.}(2008)\citenamefont{Watanabe, Yokoyama,
  Tanaka, and Inoue}}]{watanabe2008}
\bibinfo{author}{\bibfnamefont{T.}~\bibnamefont{Watanabe}},
  \bibinfo{author}{\bibfnamefont{H.}~\bibnamefont{Yokoyama}},
  \bibinfo{author}{\bibnamefont{Tanaka}}, \bibnamefont{and}
  \bibinfo{author}{\bibfnamefont{J.}~\bibnamefont{Inoue}},
  \bibinfo{journal}{Phys. Rev. B} \textbf{\bibinfo{volume}{77}},
  \bibinfo{pages}{214505} (\bibinfo{year}{2008}).

\bibitem[{\citenamefont{Ohashi et~al.}(2008)\citenamefont{Ohashi, Momoi,
  Tsnunetsugu, and Kawakami}}]{ohashi2008}
\bibinfo{author}{\bibfnamefont{T.}~\bibnamefont{Ohashi}},
  \bibinfo{author}{\bibfnamefont{T.}~\bibnamefont{Momoi}},
  \bibinfo{author}{\bibfnamefont{K.}~\bibnamefont{Tsnunetsugu}},
  \bibnamefont{and} \bibinfo{author}{\bibfnamefont{N.}~\bibnamefont{Kawakami}},
  \bibinfo{journal}{Phys. Rev. Lett.} \textbf{\bibinfo{volume}{100}},
  \bibinfo{pages}{076402} (\bibinfo{year}{2008}).

\bibitem[{\citenamefont{Sasaki et~al.}(2008)\citenamefont{Sasaki, Yoneyama, and
  Kobayashi}}]{sasaki2008}
\bibinfo{author}{\bibfnamefont{T.}~\bibnamefont{Sasaki}},
  \bibinfo{author}{\bibfnamefont{N.}~\bibnamefont{Yoneyama}}, \bibnamefont{and}
  \bibinfo{author}{\bibfnamefont{N.}~\bibnamefont{Kobayashi}},
  \bibinfo{journal}{Phys. Rev. B} \textbf{\bibinfo{volume}{77}},
  \bibinfo{pages}{054505} (\bibinfo{year}{2008}).

\bibitem[{\citenamefont{Nevidomskyy et~al.}(2008)\citenamefont{Nevidomskyy,
  Scheiber, Senechal, and Tremblay}}]{nevidomskyy2008}
\bibinfo{author}{\bibfnamefont{A.~H.} \bibnamefont{Nevidomskyy}},
  \bibinfo{author}{\bibfnamefont{C.}~\bibnamefont{Scheiber}},
  \bibinfo{author}{\bibfnamefont{D.}~\bibnamefont{Senechal}}, \bibnamefont{and}
  \bibinfo{author}{\bibfnamefont{A.-M.} \bibnamefont{Tremblay}},
  \bibinfo{journal}{Phys. Rev. B} \textbf{\bibinfo{volume}{77}},
  \bibinfo{pages}{064427} (\bibinfo{year}{2008}).

\bibitem[{\citenamefont{McKenzie}(1997)}]{mckenzie1997}
\bibinfo{author}{\bibfnamefont{R.}~\bibnamefont{McKenzie}},
  \bibinfo{journal}{Science} \textbf{\bibinfo{volume}{278}},
  \bibinfo{pages}{820} (\bibinfo{year}{1997}).

\bibitem[{\citenamefont{Lee et~al.}(2006)\citenamefont{Lee, Nagaosa, and
  Wen}}]{lee2006}
\bibinfo{author}{\bibfnamefont{P.}~\bibnamefont{Lee}},
  \bibinfo{author}{\bibfnamefont{N.}~\bibnamefont{Nagaosa}}, \bibnamefont{and}
  \bibinfo{author}{\bibfnamefont{X.-G.} \bibnamefont{Wen}},
  \bibinfo{journal}{Rev. Mod. Phys.} \textbf{\bibinfo{volume}{78}},
  \bibinfo{pages}{17} (\bibinfo{year}{2006}).

\bibitem[{\citenamefont{Hegger et~al.}(2000)\citenamefont{Hegger, Petrovic,
  Moshopoulou, Hundley, Sarrao, Fisk, and Thompson}}]{hegger2000}
\bibinfo{author}{\bibfnamefont{H.}~\bibnamefont{Hegger}},
  \bibinfo{author}{\bibfnamefont{C.}~\bibnamefont{Petrovic}},
  \bibinfo{author}{\bibfnamefont{E.~G.} \bibnamefont{Moshopoulou}},
  \bibinfo{author}{\bibfnamefont{M.~F.} \bibnamefont{Hundley}},
  \bibinfo{author}{\bibfnamefont{J.~L.} \bibnamefont{Sarrao}},
  \bibinfo{author}{\bibfnamefont{Z.}~\bibnamefont{Fisk}}, \bibnamefont{and}
  \bibinfo{author}{\bibfnamefont{J.~D.} \bibnamefont{Thompson}},
  \bibinfo{journal}{Phys. Rev. Lett.} \textbf{\bibinfo{volume}{84}},
  \bibinfo{pages}{4986} (\bibinfo{year}{2000}).

\bibitem[{\citenamefont{Hubbard}(1963)}]{hubbard1963}
\bibinfo{author}{\bibfnamefont{J.}~\bibnamefont{Hubbard}},
  \bibinfo{journal}{Proc. R. Soc. A} \textbf{\bibinfo{volume}{276}},
  \bibinfo{pages}{238} (\bibinfo{year}{1963}).

\bibitem[{\citenamefont{Kanamori}(1963)}]{kanamori1963}
\bibinfo{author}{\bibfnamefont{J.}~\bibnamefont{Kanamori}},
  \bibinfo{journal}{Prog. Theor. Phys.} \textbf{\bibinfo{volume}{30}},
  \bibinfo{pages}{275} (\bibinfo{year}{1963}).

\bibitem[{\citenamefont{Gutzwiller}(1963)}]{gutzwiller1963}
\bibinfo{author}{\bibfnamefont{M.~C.} \bibnamefont{Gutzwiller}},
  \bibinfo{journal}{Phys. Rev. Lett.} \textbf{\bibinfo{volume}{10}},
  \bibinfo{pages}{159} (\bibinfo{year}{1963}).

\bibitem[{\citenamefont{Georges et~al.}(1996)\citenamefont{Georges, Kotliar,
  Krauth, and Rozenberg}}]{georges1996}
\bibinfo{author}{\bibfnamefont{A.}~\bibnamefont{Georges}},
  \bibinfo{author}{\bibfnamefont{G.}~\bibnamefont{Kotliar}},
  \bibinfo{author}{\bibfnamefont{W.}~\bibnamefont{Krauth}}, \bibnamefont{and}
  \bibinfo{author}{\bibfnamefont{M.~J.} \bibnamefont{Rozenberg}},
  \bibinfo{journal}{Rev. Mod. Phys.} \textbf{\bibinfo{volume}{68}},
  \bibinfo{pages}{13} (\bibinfo{year}{1996}).

\bibitem[{\citenamefont{Wilson}(1975)}]{wilson1975}
\bibinfo{author}{\bibfnamefont{K.~G.} \bibnamefont{Wilson}},
  \bibinfo{journal}{Rev. Mod. Phys.} \textbf{\bibinfo{volume}{47}},
  \bibinfo{pages}{773} (\bibinfo{year}{1975}).

\bibitem[{\citenamefont{Bulla et~al.}(2008)\citenamefont{Bulla, Costi, and
  Pruschke}}]{bulla2008}
\bibinfo{author}{\bibfnamefont{R.}~\bibnamefont{Bulla}},
  \bibinfo{author}{\bibfnamefont{T.~A.} \bibnamefont{Costi}}, \bibnamefont{and}
  \bibinfo{author}{\bibfnamefont{T.}~\bibnamefont{Pruschke}},
  \bibinfo{journal}{Rev. Mod. Phys.} \textbf{\bibinfo{volume}{80}},
  \bibinfo{pages}{395} (\bibinfo{year}{2008}).

\bibitem[{\citenamefont{Kollar et~al.}(2005)\citenamefont{Kollar, Eckstein,
  Byczuk, Blümer, van Dongen, de~Cuba, Metzner, Tanaskovic, Dobrosavljevic,
  Kotliar et~al.}}]{kollar2005}
\bibinfo{author}{\bibfnamefont{M.}~\bibnamefont{Kollar}},
  \bibinfo{author}{\bibfnamefont{M.}~\bibnamefont{Eckstein}},
  \bibinfo{author}{\bibfnamefont{K.}~\bibnamefont{Byczuk}},
  \bibinfo{author}{\bibfnamefont{N.}~\bibnamefont{Blümer}},
  \bibinfo{author}{\bibfnamefont{P.}~\bibnamefont{van Dongen}},
  \bibinfo{author}{\bibfnamefont{M.~H.~R.} \bibnamefont{de~Cuba}},
  \bibinfo{author}{\bibfnamefont{W.}~\bibnamefont{Metzner}},
  \bibinfo{author}{\bibfnamefont{D.}~\bibnamefont{Tanaskovic}},
  \bibinfo{author}{\bibfnamefont{V.}~\bibnamefont{Dobrosavljevic}},
  \bibinfo{author}{\bibfnamefont{G.}~\bibnamefont{Kotliar}},
  \bibnamefont{et~al.}, \bibinfo{journal}{Ann. Phys.}
  \textbf{\bibinfo{volume}{14}}, \bibinfo{pages}{642} (\bibinfo{year}{2005}).

\bibitem[{\citenamefont{Eckstein et~al.}(2005)\citenamefont{Eckstein, Kollar,
  Byczuk, and Vollhardt}}]{eckstein2005}
\bibinfo{author}{\bibfnamefont{M.}~\bibnamefont{Eckstein}},
  \bibinfo{author}{\bibfnamefont{M.}~\bibnamefont{Kollar}},
  \bibinfo{author}{\bibfnamefont{K.}~\bibnamefont{Byczuk}}, \bibnamefont{and}
  \bibinfo{author}{\bibfnamefont{D.}~\bibnamefont{Vollhardt}},
  \bibinfo{journal}{Phys. Rev. B} \textbf{\bibinfo{volume}{71}},
  \bibinfo{pages}{235119} (\bibinfo{year}{2005}).

\bibitem[{\citenamefont{Pruschke}(2005)}]{pruschke2005}
\bibinfo{author}{\bibfnamefont{T.}~\bibnamefont{Pruschke}},
  \bibinfo{journal}{Prog. Theo. Phys. Suppl.} \textbf{\bibinfo{volume}{160}},
  \bibinfo{pages}{274} (\bibinfo{year}{2005}).

\bibitem[{\citenamefont{Rozenberg et~al.}(1995)\citenamefont{Rozenberg,
  Kotliar, Kajueter, Thomas, Rapkine, Honig, and Metcalf}}]{rozenberg1995}
\bibinfo{author}{\bibfnamefont{M.~J.} \bibnamefont{Rozenberg}},
  \bibinfo{author}{\bibfnamefont{G.}~\bibnamefont{Kotliar}},
  \bibinfo{author}{\bibfnamefont{H.}~\bibnamefont{Kajueter}},
  \bibinfo{author}{\bibfnamefont{G.~A.} \bibnamefont{Thomas}},
  \bibinfo{author}{\bibfnamefont{D.~H.} \bibnamefont{Rapkine}},
  \bibinfo{author}{\bibfnamefont{J.~M.} \bibnamefont{Honig}}, \bibnamefont{and}
  \bibinfo{author}{\bibfnamefont{P.}~\bibnamefont{Metcalf}},
  \bibinfo{journal}{Phys. Rev. Lett.} \textbf{\bibinfo{volume}{75}},
  \bibinfo{pages}{105} (\bibinfo{year}{1995}).

\bibitem[{\citenamefont{Duffy and Moreo}(1997)}]{duffy1997}
\bibinfo{author}{\bibfnamefont{D.}~\bibnamefont{Duffy}} \bibnamefont{and}
  \bibinfo{author}{\bibfnamefont{A.}~\bibnamefont{Moreo}},
  \bibinfo{journal}{Phys. Rev. B} \textbf{\bibinfo{volume}{55}},
  \bibinfo{pages}{R676} (\bibinfo{year}{1997}).

\bibitem[{\citenamefont{Hofstetter and Vollhardt}(1998)}]{hofstetter1998}
\bibinfo{author}{\bibfnamefont{W.}~\bibnamefont{Hofstetter}} \bibnamefont{and}
  \bibinfo{author}{\bibfnamefont{D.}~\bibnamefont{Vollhardt}},
  \bibinfo{journal}{Ann. Physik} \textbf{\bibinfo{volume}{7}},
  \bibinfo{pages}{48} (\bibinfo{year}{1998}).

\bibitem[{\citenamefont{Chitra and Kotliar}(1999)}]{chitra1999}
\bibinfo{author}{\bibfnamefont{R.}~\bibnamefont{Chitra}} \bibnamefont{and}
  \bibinfo{author}{\bibfnamefont{G.}~\bibnamefont{Kotliar}},
  \bibinfo{journal}{Phys. Rev. Lett.} \textbf{\bibinfo{volume}{83}},
  \bibinfo{pages}{2386} (\bibinfo{year}{1999}).

\bibitem[{\citenamefont{Zitzler et~al.}(2004)\citenamefont{Zitzler, Tong,
  Pruschke, and Bulla}}]{zitzler2004}
\bibinfo{author}{\bibfnamefont{R.}~\bibnamefont{Zitzler}},
  \bibinfo{author}{\bibfnamefont{N.-H.} \bibnamefont{Tong}},
  \bibinfo{author}{\bibfnamefont{T.}~\bibnamefont{Pruschke}}, \bibnamefont{and}
  \bibinfo{author}{\bibfnamefont{R.}~\bibnamefont{Bulla}},
  \bibinfo{journal}{Phys. Rev. Lett.} \textbf{\bibinfo{volume}{93}},
  \bibinfo{pages}{016406} (\bibinfo{year}{2004}).

\bibitem[{\citenamefont{van Dongen}(1991)}]{dongen1991}
\bibinfo{author}{\bibfnamefont{P.~G.~J.} \bibnamefont{van Dongen}},
  \bibinfo{journal}{Phys. Rev. Lett.} \textbf{\bibinfo{volume}{67}},
  \bibinfo{pages}{757} (\bibinfo{year}{1991}).

\bibitem[{\citenamefont{Eckstein et~al.}(2007)\citenamefont{Eckstein, Kollar,
  Potthoff, and Vollhardt}}]{eckstein2007}
\bibinfo{author}{\bibfnamefont{M.}~\bibnamefont{Eckstein}},
  \bibinfo{author}{\bibfnamefont{M.}~\bibnamefont{Kollar}},
  \bibinfo{author}{\bibfnamefont{M.}~\bibnamefont{Potthoff}}, \bibnamefont{and}
  \bibinfo{author}{\bibfnamefont{D.}~\bibnamefont{Vollhardt}},
  \bibinfo{journal}{Phys. Rev. B} \textbf{\bibinfo{volume}{75}},
  \bibinfo{pages}{125103} (\bibinfo{year}{2007}).

\bibitem[{\citenamefont{Potthoff}(2003)}]{potthoff2003}
\bibinfo{author}{\bibfnamefont{M.}~\bibnamefont{Potthoff}},
  \bibinfo{journal}{Eur. Phys. J. B} \textbf{\bibinfo{volume}{32}},
  \bibinfo{pages}{429} (\bibinfo{year}{2003}).

\bibitem[{\citenamefont{Fleck et~al.}(1999)\citenamefont{Fleck, Lichtenstein,
  Oles, and Hedin}}]{fleck1999}
\bibinfo{author}{\bibfnamefont{M.}~\bibnamefont{Fleck}},
  \bibinfo{author}{\bibfnamefont{A.~I.} \bibnamefont{Lichtenstein}},
  \bibinfo{author}{\bibfnamefont{A.~M.} \bibnamefont{Oles}}, \bibnamefont{and}
  \bibinfo{author}{\bibfnamefont{L.}~\bibnamefont{Hedin}},
  \bibinfo{journal}{Phys. Rev. B} \textbf{\bibinfo{volume}{60}},
  \bibinfo{pages}{5224} (\bibinfo{year}{1999}).

\bibitem[{\citenamefont{Aspelmeier}(2008)\citenamefont{Aspelmeier}}]{Aspelmeier}
\bibinfo{author}{\bibfnamefont{T.}~\bibnamefont{Aspelmeier}},
  \bibinfo{journal}{private communication}. 
\end{thebibliography}
\end{document}